\documentclass[12pt,draftclsnofoot,onecolumn]{IEEEtran}
\IEEEoverridecommandlockouts
\usepackage[utf8]{inputenc}
\usepackage{bm}
\usepackage{amsfonts}
\usepackage{amssymb}
\usepackage{amsmath}
\usepackage{algorithm}
\usepackage{algpseudocode}
\usepackage{graphicx}
\usepackage{siunitx}
\usepackage{cite}
\usepackage[font=small]{caption}
\usepackage{multicol}

\title{Sensor Selection and Distributed Quantization for Energy Efficiency in Massive MTC}
\author{Sergi~Liesegang,~\IEEEmembership{Student Member,~IEEE}, Olga Mu\~noz,~\IEEEmembership{Member,~IEEE} and Antonio Pascual-Iserte,~\IEEEmembership{Senior Member,~IEEE}
\thanks{The work presented in this paper has been carried out within the framework of the project ROUTE56 (PID2019-104945GB-I00 / AEI / 10.13039/501100011033) funded by the Agencia Estatal de Investigaci\'on (Spanish Ministry of Science and Innovation); the FPI grant BES-2017-079994, funded by the Spanish Ministry of Science, Innovation and Universities; and the grant 2017 SGR 578, funded by the Catalan Government (AGAUR, Departament de Rercerca i Universitats, Generalitat de Catalunya).}
\thanks{The authors are with the Department of Signal Theory and Communications,
Universitat Polit\`ecnica de Catalunya, Barcelona 08034, Spain (e-mail: sergi.liesegang@upc.edu, olga.munoz@upc.edu, antonio.pascual@upc.edu).

\copyright 2021 IEEE. Personal use of this material is permitted. Permission from IEEE must be obtained for all other uses, in any current or future media, including reprinting/republishing this material for advertising or promotional purposes, creating new collective works, for resale or redistribution to servers or lists, or reuse of any copyrighted component of this work in other works.

DOI: 10.1109/TCOMM.2021.3112206

}}

\begin{document}

\markboth{Accepted Paper at IEEE Transactions on Communications (vol. 69, no. 12, Dec. 2021)}
{}

\maketitle

\begin{abstract}
This paper presents an estimation approach within the framework of uplink massive machine-type communications (mMTC) that considers the energy limitations of the devices. We focus on a scenario where a group of sensors observe a set of parameters and send the measured information to a collector node (CN). The CN is responsible for estimating the original observations, which are spatially correlated and corrupted by measurement and quantization noise. Given the use of Gaussian sources, the minimum mean squared error (MSE) estimation is employed and, when considering temporal evolution, the use of Kalman filters is studied. Based on that, we propose a device selection strategy to reduce the number of active sensors and a quantization scheme with adjustable number of bits to minimize the overall payload. The set of selected sensors and quantization levels are, thus, designed to minimize the MSE. For a more realistic analysis, communication errors are also included by averaging the MSE over the error decoding probabilities. We evaluate the performance of our strategy in a practical mMTC system with synthetic and real databases. Simulation results show that the optimization of the payload and the set of active devices can reduce the power consumption without compromising the estimation accuracy.
\end{abstract}


\section{Introduction} \label{sec:1}
Machine-type communications (MTC) play an essential role in the evolution of future mobile systems \cite{Che17}. They constitute a type of network where a set of devices communicate without human supervision and where the number of connected terminals is expected to grow exponentially in the next decade \cite{Wan17}. 3GPP standards such as enhanced MTC, also known as long-term-evolution for machines, and narrow-band Internet-of-Things (IoT), are only two examples of the impact of MTC on cellular communications \cite{Els17}. In many MTC applications, as those under the umbrella of the IoT, the devices involved can be power constrained, specially if the replacement or charging of batteries is difficult \cite{Raz17}. That is why high energy efficiency is pursued and is one of the key elements in this kind of networks \cite{Boc16,Sha20}.

In this work, we will focus on an uplink (UL) scenario where a serving base station or collector node (CN) estimates a set of parameters based on the noisy measurements (or observations) provided by a large number of sensors. This setup is referred to as massive MTC (mMTC). In mMTC, according to the type of traffic, we can distinguish two main classes: event-driven (e.g., emergency scenarios) or periodic (e.g., smart metering, industrial control networks, among others) \cite{Boc16,3GPP45820}. In the first case, due to the low or sporadic traffic arrival patterns, the use of random-access protocols presents clear benefits \cite{Wee21}. Contrarily, for applications characterized by periodic patterns (e.g., the information is retrieved by polling the sensors every certain time) or deterministic arrivals (e.g., data is communicated within a given time frame), conflict-free access mechanisms such as configured scheduling, become more suitable than random-access strategies \cite{3GPP38912,Jur19}. This paper focuses on this last case and, as in \cite{Jur19}, considers a time-division multiplexing scheduling mechanism.

Given the high spatial density of mMTC deployments, the data sensed by different devices can be significantly correlated \cite{Sha13}, which can be exploited to improve the accuracy of the estimation. Hence, a natural question is how to manipulate all the gathered information to reduce the overall payload and power consumption of the sensors with still good estimation accuracy.

Parameter estimation has been widely studied in the literature and the optimal strategy depends on the scenario under study \cite{Zhu05,Li09,Tal14}. In the case of Gaussian sources, it is common to consider the minimum squared error (MSE) estimation as the optimum approach \cite{Kay93}. However, in most works, transmission errors are not taken into account, and a noiseless channel is usually considered \cite{Fan10,Zha19}. As these errors may affect significantly the performance of the system, in this paper we average the MSE over the different error decoding probabilities.

Based on that, we will consider a communication scheme in which only a subset of sensors will be selected to transmit their observations (cf. \cite{Zhu05}). The decision concerning which sensors remain active will be driven by the objective of minimizing the MSE of the estimate calculated by the CN. Thereby, the degree of correlation between the measurements will have an impact on the selection of sensors. In short, sensors with highly correlated data will not transmit all their measurements, but some of them will remain silent. This way, we will reduce the power consumption within the network. Note that for a Gaussian source, the correlation between the sensors information is captured by the covariance matrix of the observations \cite{Sca05}.

Moreover, we will consider that sensors quantize their measurements, which will also help to reduce the network traffic and, therefore, the overall power consumption \cite{Stu16}. In this setup, we will consider the use of uniform scalar quantizers, for which we will characterize the resulting quantization noise and study its impact on the MSE  \cite{Dra09}. In that sense, the different precision levels in each device (i.e., the number of quantization bits) will be designed to minimize the MSE, leveraging the correlation in the sensed data. 

Finally, in the cases where the observed parameters can vary over time (e.g., temperature during the day), the sensor measurements might also be temporally correlated \cite{Vur04}. Thereby, to take into account these situations, we will extend our study to the case of systems with memory. In that sense, we will use first-order Markov processes to model these dynamics and, since the goal is to minimize the MSE, we will consider the use of Kalman filters to produce the estimates at the CN \cite{Mse08,Fre19}. Accordingly, the device selection and the number of quantization bits will be optimized over time and the resulting MSE evolution will be analyzed.

\subsection{Prior Work} \label{sec:1.1}
Based on the fundamentals derived by S. M. Kay in 1993 \cite{Kay93}, early works like \cite{Xia05,Zhu05,Fan09} tackle the problem of distributed estimation in a scenario where a set of devices send their spatially correlated data to a fusion center through a noiseless channel. In particular, considering Gaussian sources, the authors in \cite{Xia05} derive the optimal minimum MSE (MMSE) estimator when the measurement noise is inhomogeneous (i.e., not equally distributed for all sensors). In \cite{Zhu05}, the authors extend these derivations to the case of temporally correlated data (with noiseless channels). The case of power constrained devices (and no transmission errors) is analyzed in \cite{Fan09}, where the authors also consider the presence of collaborative clusters. In the recent work \cite{Zha19}, the authors design an estimation strategy based on distributed compression and dimensionality reduction to comply with the bandwidth constraints in an error-free communication environment.

Within this framework, the optimal selection of the sensing devices has also been extensively pursued (cf. \cite{Jos09}). In fact, works \cite{Zhu05,Fan09} already introduce this problem by reducing the message dimension. On the contrary, the authors in \cite{Mse12} propose the use of data censoring for reducing the information to be sent (which can be seen as an alternative to sensor selection). Besides, with the help of kernel regression, a fast and low-complexity approach based on matrix completion and extrapolation is studied in \cite{Gim19}, which accounts for prior information in the data. However, these works consider negligible transmission errors, which is an unrealistic assumption in mMTC networks (where communication resources are usually limited).

Finally, given their relevance in practical systems, data quantizers have also been considered in some of these works (e.g., \cite{Mse12}). The reason is that efficient compression strategies can be obtained when optimizing the quantization stage \cite{Xia06}. For instance, the authors in \cite{Fan08b} derive the set of optimal quantizers when only $1$-bit of resolution is available. Besides, in \cite{Fre19}, the authors study the potential of analog mappings in a fading multiple-access environment and optimize the precision levels of the quantizers. Another example can be found in \cite{Nik17}, where the authors rely on encoded sensing to partition the network into groups of sensors that jointly encode and transmit their information to the sink. Nevertheless, most of these works do not take into account the power limitations of the devices involved (which is crucial in current mMTC deployments).

\subsection{Contributions} \label{sec:1.2}
The contributions of this paper are listed in what follows:
\begin{itemize}
    \item[1)] We introduce a distributed estimation scheme based on sensor selection and uniform scalar quantization with different levels of precision. Our approach is firstly formulated for a memoryless mMTC system and benefits from the spatial correlation in the sensed data.
    \item[2)] We propose a parameter estimation based on the MMSE criterion. The MMSE estimate is derived considering actual data transmission and communication errors. The analytic closed-form expression for the MSE is characterized and a tractable upper bound is proposed.
    \item[3)] We design an iterative algorithm to optimize the set of selected devices and the number of quantization bits, taking into consideration the power consumption within the network. Two strategies are presented: separate (alternate) and joint optimization.
    \item[4)] We study the extension to temporally correlated data based on the use of the Kalman filter. Accordingly, the resulting MSE and the MMSE estimate are derived.
    \item[5)] We present numerical simulations for synthetically generated data and real data. In both cases, results highlight the performance of our approach and justify the selection and quantization of measurements in mMTC deployments.
\end{itemize}

\subsection{Organization} \label{sec:1.3}
The remainder of this work is structured as follows. In Section~\ref{sec:2}, the system model and the estimation scheme (i.e., sensor selection and distributed quantization) are described. In Section~\ref{sec:3}, the parameter estimation and the MSE are characterized. In Section~\ref{sec:4}, the optimization problem is formulated and the two approaches to find a feasible solution are proposed. In Section~\ref{sec:5}, the extension to the system with memory is discussed. Numerical simulations are shown in Section~\ref{sec:6} for synthetic and real data and, finally, conclusions are presented in Section~\ref{sec:7}.

\subsection{Notation} \label{sec:1.4}
In this work, scalars are denoted by italic letters. Boldface lower-case and upper-case letters denote vectors and matrices, respectively. The transpose, inverse and trace operators are denoted by $(\cdot)^{\textrm{T}}$, $(\cdot)^{\textrm{-1}}$ and $\textrm{tr}(\cdot)$, respectively. The expectation operator is denoted by $\mathbb{E}[\cdot]$. $\mathbf{I}_m$ denotes the identity matrix of size $m \times m$ and $\bm{1}_m$ denotes the all ones column vector of length $m$. For a given set $\mathcal{A}$, the cardinality is denoted by $\vert \mathcal{A} \vert$ and $\emptyset$ denotes the empty set. $\mathbb{R}^{m \times n}$ and $\mathbb{N}^{m \times n}$ denote the $m$ by $n$ dimensional real space and natural space, respectively. The multivariate Gaussian distribution with mean $\bm{\mu}$ and covariance matrix $\bm{C}$ is denoted by $\mathcal{N}(\bm{\mu},\bm{C})$.

\section{System Model} \label{sec:2}
\begin{figure}[t]
\centering
\includegraphics[scale = 0.5]{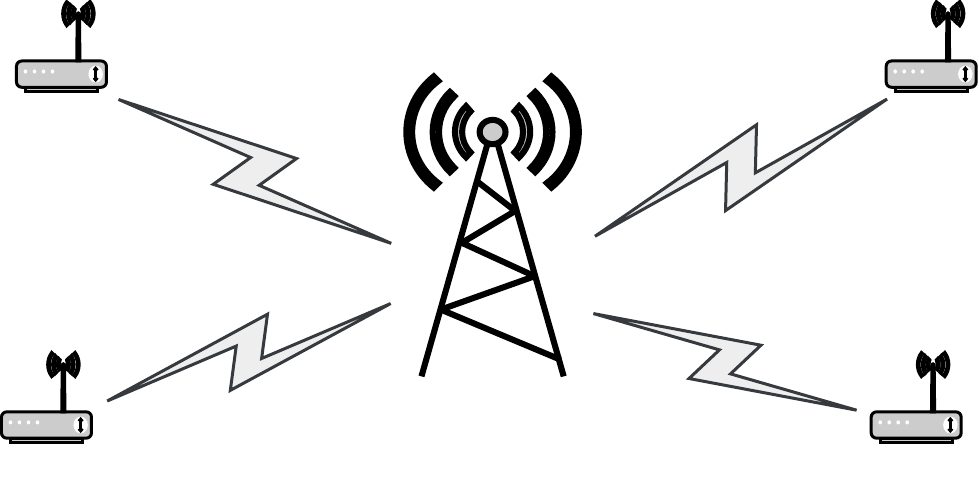}
\caption{Illustrative scenario with $M = 4$.}
\label{fig:1}
\end{figure}

Throughout this work, we consider a scenario with a set of $M$ sensors connected to a CN. An illustrative example with $M = 4$ is depicted in Fig.~\ref{fig:1}. Each of these sensors measures a noisy version $x_i$ of a parameter $\theta_i$, with $i \in \{1,\ldots,M\}$. More specifically, in vector notation we have \cite{Kho20}: 
\begin{equation}
    \bm{x} = \bm{\theta} + \bm{w} \in \mathbb{R}^M,
    \label{eq:1}
\end{equation}
where $\bm{x} = [x_1, \ldots, x_M]^{\textrm{T}}$ is the set of observations of the (physically separated) devices, $\bm{\theta} = [\theta_1, \ldots, \theta_M]^{\textrm{T}} \in \mathbb{R}^M$ is the vector containing the different (but correlated) parameters, and $\bm{w} \in \mathbb{R}^M$ is the corresponding measurement noise vector. Each element in $\bm{x}$ is measured independently by each individual sensor. We assume that the parameter vector is a realization of a Gaussian process with mean $\bm{\mu}_{\bm{\theta}} \in \mathbb{R}^M$ and covariance matrix $\bm{C}_{\bm{\theta}} \in \mathbb{R}^{M \times M}$, i.e., $\bm{\theta} \sim \mathcal{N}\left(\bm{\mu}_{\bm{\theta}},\bm{C}_{\bm{\theta}}\right)$. Both mean and covariance are assumed to be known. In addition, we also consider the noise vector $\bm{w}$ to be independent of $\bm{\theta}$ and Gaussian distributed with zero mean and covariance matrix $\bm{C}_{\bm{w}} \in \mathbb{R}^{M \times M}$, i.e., $\bm{w} \sim \mathcal{N}\left(\bm{0},\bm{C}_{\bm{w}}\right)$. Note that this model also includes, as a particular case, the scenario where all sensors measure the same parameter ($\theta_i = \theta_j$  $\forall i,j$), i.e., $\bm{\mu}_{\bm{\theta}} = \mu_{\theta}\bm{1}_M$ and $\bm{C}_{\bm{\theta}}=\sigma^2\bm{1}_M\bm{1}_M^{\textrm{T}}$ (where $\sigma$ is the standard deviation).

The elements of the covariance matrix $\bm{C}_{\bm{\theta}}$ can be modeled, for example, as \cite{Lie20}
\begin{equation}
    [\bm{C}_{\bm{\theta}}]_{i,j} = \sigma_{i} \sigma_j K(d_{i,j}), \quad 1 \leq i \leq M, \quad 1 \leq j \leq M,
    \label{eq:2}
\end{equation}
where $\sigma_i$ is the standard deviation of the parameter $\theta_i$ measured by sensor $i$. The term $K(d_{i,j})$ is the correlation factor between the parameters of sensors $i$ and $j$ with $\vert K(d_{i,j}) \vert \leq 1$, and that, usually, depends on the distance between them $d_{i,j}$. Different models can be adopted for the function $K(d_{i,j})$. As an example, a well-known and accepted model for sensor networks is an exponential distance (cf. \cite{Vur04,Nik17}). However, to keep our analysis general, the expression of $K(d_{i,j})$ will remain unspecified until the simulations section.

As mentioned before, sensors will quantize their measurements according to different precision levels to reduce the payload and power consumption. In this setup, we consider uniform scalar quantizers with step size $\Delta_i$. Note that, since the different scalar observations are available at different sensors positioned at separate locations, vector quantization techniques cannot be applied. Thereby, each sensor will use $n_i$ bits of precision and, thus, the quantization step yields
\begin{equation}
    \Delta_i = S_i/2^{n_i},
    \label{eq:3}
\end{equation}
where $S_i$ is the dynamic margin of the quantizer. However, given the assumption of Gaussian random variables (RVs), the support of the measurement $x_i$ is not bounded. That is why we will consider $S_i$ to be $6$ times the standard deviation of $x_i$ since $99.73\%$ of the values lie within that interval. For a white noise with covariance matrix $\bm{C}_{\bm{w}} = \sigma_w^2 \mathbf{I}_M$, we would have $S_i \approx 6 \sqrt{\sigma_i^2 + \sigma_w^2}$.

As a result, the information to be transmitted by each sensor reads as \cite{Wid08}
\begin{equation}
    y_i = q_i(x_i) = x_i + \eta_i \in \mathbb{R}, \quad 1 \leq i \leq M,
    \label{eq:4}
\end{equation}
where $q_i(\cdot)$ represents the quantization function and $\eta_i$ is the corresponding quantization noise. 

Note that, for a sufficiently small step size $\Delta_i$, this noise can be shown to be uncorrelated with $x_i$, and that the first and second order moments can be safely approximated by those of a noise uniformly distributed in $[-\Delta_i/2,\Delta_i/2]$. In fact, for Gaussian RVs, this approximation holds for values of $\Delta_i$ smaller than the standard deviation of $x_i$ \cite{Wid08}.

In order to avoid unnecessary power consumption, only a subset of sensors will be selected to transmit their information to the CN. Thereby, the chosen sensors will transmit their quantized observation to the CN through separate orthogonal channels using different modulation and coding schemes (MCSs). More specifically, following the 5G standard \cite{Sha17}, we will adopt a communication strategy in which time is divided into frames of $N$ slots, each of duration $T_s$ seconds and bandwidth $B_s$ Hertz \cite{Boc18,Cai18}. Both parameters are assumed to be fixed and known. Accordingly, each sensor will be allocated a single slot and, thus, no collisions nor interference will be experienced (cf. \cite{Jur19}). Hence, with this approach, only $N$ sensors are allowed to transmit during each frame. Recall that we focus on applications where sensor information is generated periodically (and will be transmitted if selected by the CN). In these scenarios, it is reasonable to assume that the CN has information regarding how many devices are sensing. Note that this assumption is valid for this type of sensor-based mMTC network, but it does not hold for systems with unpredictable or random traffic patterns.

Let $\bm{z} = [z_1, \ldots, z_N]^{\textrm{T}} \in \mathbb{R}^N$ denote the corresponding set of $N$ measurements from the selected sensors. This vector can be expressed in terms of the following binary selection matrix: 
\begin{equation}
    \bm{V} = [ \bm{v}_1^{\textrm{T}}, \ldots, \bm{v}_N^{\textrm{T}} ]^{\textrm{T}} \in \{0,1\}^{N \times M},
    \label{eq:5}
\end{equation}
where each row $\bm{v}_i \in \{0,1\}^{1 \times M}$, with $i \in \{1,\ldots,N\}$, is a unit vector indicating which sensor is transmitting (i.e., the position of the non-zero element determines the active sensing device). In this work, we consider that the rest of the sensors remain silent and discard their information. 

As an illustrative example, let us consider $M = 5$, $N = 3$ and
\begin{equation}
\bm{V} = \begin{bmatrix} 1 & 0 & 0 & 0 & 0 \\ 0 & 0 & 1 & 0 & 0 \\ 0 & 0 & 0 & 1 & 0 \end{bmatrix}.
\label{eq:6}
\end{equation}
In this case, the first, third, and fourth sensor are selected to transmit their measurements. Note that all unit rows in $\bm{V}$ must be different, which is equivalent to the constraint $\bm{V}\bm{V}^{\textrm{T}} = \mathbf{I}_N$.

\begin{figure*}[t]
\centering 
\includegraphics[scale = 1]{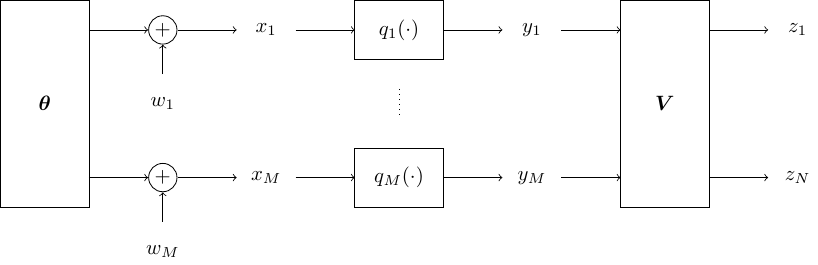} 
\caption{Estimation scheme prior to transmission.}
\label{fig:2}
\end{figure*}

The vector $\bm{z}$ can be written as
\begin{equation}
    \bm{z} = \bm{V} \bm{y} \in \mathbb{R}^N,
    \label{eq:7}
\end{equation}
where $\bm{y} = [y_1, \ldots, y_M]^{\textrm{T}}$ is the original set of quantized measurements. The resulting estimation scheme (prior to transmission) is depicted in Fig.~\ref{fig:2}.

Finally, based on the correctly received information, the CN will estimate the parameter vector $\bm{\theta}$. Thus, as mentioned before, we need to include the possible communication errors during the data transmission. This will be discussed in the upcoming section. 

\section{Parameter Estimation} \label{sec:3}
This section is devoted to characterize the MMSE estimate in the presence of communication errors. To do so, we first study the error decoding probability, which will affect the set of available observations at the CN and the resulting MSE.

\subsection{Error Decoding Probability} \label{sec:3.1}
Prior to transmission, the information (i.e., quantization) bits $n_i$ of each sensor are transformed into $n_i^{\textrm{cod}}$ coded bits and later modulated into packets of $n_i^{\textrm{sym}}$ symbols. Defining $R_i = n_i/n_i^{\textrm{cod}}$ as the coding rate (information bits/coded bits), and $L_i = n_i^{\textrm{cod}}/n_i^{\textrm{sym}}$ as the modulation size (coded bits/symbol), we have
\begin{equation}
    n_i = T_s B_s R_i L_i,
    \label{eq:8}
\end{equation}
where $T_s B_s$ represents the number of symbols per slot. Therefore, given that both $T_s$ and $B_s$ are fixed, the number of (information) bits to be transmitted $n_i$ only depends on the MCS, i.e., the different combinations of $R_i$ and $L_i$. Note that the product $R_i L_i$ yields the spectral efficiency (that is, the ratio between number of information bits and the number of transmitted symbols).

With the above considerations, the CN will decode the (selected) quantized observations with individual error probabilities, which depend on the channel quality, the MCS, and the number of bits $n_i$ of each sensor. In particular, the (coded) packet error rate (PER) corresponding to all the bits transmitted by sensor $i$ in a single slot can be defined as
\begin{equation}
    \textrm{PER}_i = \mathcal{C}_i\left(\textrm{PER}_i^{\textrm{raw}}\right),
    \label{eq:9}
\end{equation}
where $\textrm{PER}_i^{\textrm{raw}} \geq \textrm{PER}_i$ is the (raw) PER computed at the output of the channel and before the channel decoder for the set of $n_i^{\textrm{cod}}$ coded bits. Accordingly, $\mathcal{C}_i(\cdot)$ is the function that relates both probabilities and that depends exclusively on the coding rate $R_i$ and the channel encoder \cite{Pro08}. For the sake of generality, the functions $\mathcal{C}_i(\cdot)$ will not be detailed until Section~\ref{sec:6}.

In the case of $\textrm{PER}_i^{\textrm{raw}}$, assuming independent bit errors in the communication (due to channel and noise effects)\footnote{In the case of slowly varying channels, this condition can be easily met with the help of bit interleavers \cite{Tse05}.}, the expression can be written as
\begin{equation}
\textrm{PER}_i^{\textrm{raw}} = 1 - (1 - \textrm{BER}_i^{\textrm{raw}})^{n_i^{\textrm{cod}}} \stackrel{(a)}{=}  1 - (1 - \textrm{BER}_i^{\textrm{raw}})^{T_s B_sL_i},
\label{eq:10}
\end{equation}
where (a) follows from equation \eqref{eq:8} and the definition of $R_i$. 
The term $\textrm{BER}_i^{\textrm{raw}}$ is the (raw) bit error rate, which is independent of the channel code, and only depends on the modulation scheme and the signal-to-noise ratio (SNR) of the received signal from sensor $i$, namely
\begin{equation}
    \rho_i = \frac{G_i P_i}{B_s N_o},
    \label{eq:11}
\end{equation}
where $G_i$ is the channel gain (including the path-loss) associated to the link between sensor $i$ and the CN, $P_i$ is the sensor transmit power, and $N_o$ is the noise spectral density \cite{Tse05}.

Note that, under the previous assumptions, the power consumption and power reduction can be defined as $\textrm{tr}\left(\bm{V}\bm{P} \bm{V}^{\textrm{T}}\right)$ and $\textrm{tr}\left(\bm{P}\right) - \textrm{tr}\left(\bm{V}\bm{P} \bm{V}^{\textrm{T}}\right)$, respectively, where $\bm{P} = \textrm{diag}\left(P_1,\ldots,P_M\right)$ is the diagonal matrix with all transmit powers. As we will discuss in Section~\ref{sec:4}, $\textrm{tr}\left(\bm{V}\bm{P} \bm{V}^{\textrm{T}}\right)$ will be constrained to a maximum value in order to limit the network power consumption.

\subsection{MMSE Estimate} \label{sec:3.2}
Once the transmitted bits are decoded, the CN will use them to estimate the parameter vector $\bm{\theta}$. Accordingly, the device selection and the number of quantization bits $\bm{n} = [n_1,\ldots,n_M]^{\textrm{T}}$ will be designed to optimize the estimation accuracy. 

To evaluate the whole procedure, in this work, we consider the MSE as performance indicator. As a result, a suitable estimate of $\bm{\theta}$ is the linear MMSE estimate given by (we will first assume that no errors occur in the communication link)  \cite{Zhu05}:
\begin{equation}
    \hat{\bm{\theta}} = \mathbb{E}[\bm{\theta} \vert \bm{z}] = \bm{C}_{\bm{\theta} \bm{z}}\bm{C}_{\bm{z}}^{-1}\left(\bm{z} - \mathbb{E}[\bm{z}]\right) + \bm{\mu}_{\bm{\theta}}.
    \label{eq:12}
\end{equation}

Thereby, considering the uncorrelation between $\bm{x} = [x_1,\ldots,x_M]^{\textrm{T}}$ and the quantization errors $\bm{\eta} = [\eta_1,\ldots,\eta_M]^{\textrm{T}}$, the statistical terms in \eqref{eq:12} yield
\begin{equation}
    \bm{C}_{\bm{\theta} \bm{z}} = \mathbb{E}\left[\bm{\theta}\bm{z}^{\textrm{T}}\right] = \mathbb{E}\left[\bm{\theta}(\bm{\theta} + \bm{w} + \bm{\eta})^{\textrm{T}} \bm{V}^{\textrm{T}}\right] = \bm{C}_{\bm{\theta}} \bm{V}^{\textrm{T}},
    \label{eq:13}
\end{equation}
\begin{equation}
    \bm{C}_{\bm{z}} = \mathbb{E}\left[\bm{z}\bm{z}^{\textrm{T}}\right] = \bm{V}\left(\bm{C}_{\bm{\theta}}  + \bm{C}_{\bm{w}} + \bm{C}_{\Delta}\right) \bm{V}^{\textrm{T}} ,
    \label{eq:14}
\end{equation}
\begin{equation}
    \mathbb{E}[\bm{z}] = \bm{V} \bm{\mu}_{\bm{\theta}},
    \label{eq:15}
\end{equation}
where $\bm{C}_{\Delta} = \textrm{diag}\left(\Delta_1^2/12, \ldots, \Delta_M^2/12\right)$ in \eqref{eq:14}. Note that this expression corresponds to the second-order moments of the uncorrelated and uniformly distributed quantization errors.

Consequently, the MSE is given by the trace of the error covariance matrix $\bm{\Sigma}$ \cite{Zhu05}
\begin{equation}
    \textrm{tr}\left(\bm{\Sigma}\right) = \textrm{tr}\left(\bm{C}_{\bm{\theta}} - \bm{C}_{\bm{\theta} \bm{z}} \bm{C}_{\bm{z}}^{-1} \bm{C}_{\bm{z} \bm{\theta}} \right),
    \label{eq:16}
\end{equation}
where, following the derivations in \eqref{eq:13}, it can be seen that $\bm{C}_{\bm{z} \bm{\theta}} = \bm{V}\bm{C}_{\bm{\theta}}$.

Note that, in the previous expressions, we have considered that there were no errors in the communication. However, as we discussed in the previous subsection, packets might be incorrectly decoded with probability $\textrm{PER}_i$ for the set of active sensors. In that case, that packet would be thrown away, and the corresponding elements of $\bm{z}$ would be discarded as well. In order to formulate this, we can replace the matrix $\bm{V}$ by the following:
\begin{equation}
    \bm{U}_{\mathcal{I}} = \bm{A}_{\mathcal{I}} \bm{V} \in \{0,1\}^{\vert \mathcal{I}\vert \times M},
    \label{eq:17}
\end{equation}
where $\mathcal{I}$ is the set of active sensors with correctly decoded messages and $\bm{A}_{\mathcal{I}} \in \{0,1\}^{\vert \mathcal{I} \vert \times N}$ is the corresponding flat matrix indicating the absence of errors. In that sense, similar to $\bm{V}$, the non-zero positions in the (unit) rows of matrix $\bm{A}_{\mathcal{I}}$ define the (active) sensing devices with perfectly decoded messages. This way, $\bm{U}_{\mathcal{I}}$ is a binary flat matrix that represents the information finally available at the CN (after device selection and packet decoding).  

Considering the decoding errors, now the set of available measurements at the CN yields
\begin{equation}
    \bm{z}_{\mathcal{I}} = \bm{U}_{\mathcal{I}}\bm{y} \in \mathbb{R}^{\vert \mathcal{I} \vert},
    \label{eq:18}
\end{equation}
and, thus, the corresponding MMSE estimate can be written as
\begin{equation}    
    \hat{\bm{\theta}}_{\mathcal{I}} = \mathbb{E}[\bm{\theta} \vert \bm{z}_{\mathcal{I}}] = \bm{C}_{\bm{\theta}}  \bm{U}_{\mathcal{I}}^{\textrm{T}}\left(\bm{U}_{\mathcal{I}}\left(\bm{C}_{\bm{\theta}} + \bm{C}_{\bm{w}} + \bm{C}_{\Delta}\right) \bm{U}_{\mathcal{I}}^{\textrm{T}}\right)^{-1} \left(\bm{z}_{\mathcal{I}} - \mathbb{E}[\bm{z}_{\mathcal{I}}]\right) + \bm{\mu}_{\bm{\theta}},
    \label{eq:19}    
\end{equation}
with MSE:
\begin{equation}
    \varepsilon_{\mathcal{I}} \triangleq \textrm{tr}\big(\bm{C}_{\bm{\theta}} - \bm{C}_{\bm{\theta}} \bm{U}_{\mathcal{I}}^{\textrm{T}}  \left(\bm{U}_{\mathcal{I}}\left( \bm{C}_{\bm{\theta}} + \bm{C}_{\bm{w}} + \bm{C}_{\Delta}\right)\bm{U}_{\mathcal{I}}^{\textrm{T}}\right)^{-1} \bm{U}_{\mathcal{I}}  \bm{C}_{\bm{\theta}}\big).
    \label{eq:20}
\end{equation}

Hence, we can define the overall MSE by averaging over the different error probabilities:
\begin{equation}
    \varepsilon \triangleq \sum_{k = 0}^{N}\sum_{\mathcal{I} \in \mathcal{F}_k}  \varepsilon_{\mathcal{I}} \prod_{i \in \mathcal{I}} \left(1- \textrm{PER}_i\right) \prod_{j \not\in \mathcal{I}} \textrm{PER}_j,
    \label{eq:21}
\end{equation}
where $\mathcal{F}_k$ represents all the tuples of size $N - k$ taken from the set of active sensors defined by $\bm{V}$. Thus, the subindex $k$ indicates the number of incorrect messages. Going back to the previous example (cf. \eqref{eq:6}), we have that $\mathcal{F}_1 = \{ \{1,3\},\{1,4\},\{3,4\}\}$ and $\mathcal{F}_2 = \{\{1\},\{3\},\{4\}\}$. Note that $\mathcal{F}_0 = \{\{1,3,4 \}\}$ always refers to the case of no errors.

\section{Problem Formulation and Solution} \label{sec:4}
The purpose of this work is to derive a selection strategy and a quantization scheme that minimize the MSE in \eqref{eq:21} so that only $N$ sensors are active and transmitting during a single frame with a limited power consumption. In particular, we are interested in designing the optimal selection matrix $\bm{V}$ and the optimal number of quantization bits $\bm{n}$.

Thereby, the optimization problem can be formulated as follows:
\begin{equation}    
    (P1): \{\bm{V}^{\star},\bm{n}^{\star}\} = \underset{\stackrel{\bm{V} \in \{0,1\}^{N \times M}}{ \bm{n} \in \mathbb{N}^M}}{\textrm{argmin}} \, \, \varepsilon  \quad \textrm{s.t.} \quad \bm{V} \bm{V}^{\textrm{T}} = \mathbf{I}_N,  \quad \textrm{tr}\left(\bm{V}\bm{P}\bm{V}^{\textrm{T}}\right) \leq \delta,
    \label{eq:22}
\end{equation}
where, as mentioned before, the first constraint ensures that only $N$ sensors are selected, and the second one ensures that the power consumption (of the selected devices) is below a given threshold $\delta$. Recall that the individual powers in $\bm{P}$ are considered to be fixed (not optimized) and can be different among them. This way, thanks to the second constraint, we avoid selecting simultaneously sensors with (only) high powers, yet concentrate on combining devices with high and low (or simply low) consumption. Besides, note that in realistic scenarios, the number of bits might be further restricted to minimum and maximum values, both depending on the capability of the sensing devices. Hence, the space dimension of $\bm{n}$ can be considerably reduced.

Unfortunately, given the expression of the objective function and the discrete nature of the optimization variables, the problem in \eqref{eq:22} is combinatorial and has exponential complexity. That is why, in this work, we will use some simplifications to reduce the optimization complexity and find a (feasible) sub-optimal solution with still good performance.

First, since we want to minimize $\varepsilon$, let us define the following upper bound:
\begin{equation}
    \varepsilon \leq p_{\mathcal{F}_{0}} \varepsilon_{\mathcal{F}_{0}} + \sum_{k = 1}^{K} \sum_{\mathcal{I} \in \mathcal{F}_k} p_{\mathcal{I}} \varepsilon_{\mathcal{I}} + p_{K} \varepsilon_{\mathcal{F}_N} \triangleq \bar{\varepsilon}_K,
    \label{eq:23}
\end{equation}
where
\begin{equation}
    p_{\mathcal{F}_{0}} \triangleq \prod_{i \in \mathcal{I}, \mathcal{I} \in \mathcal{F}_{0}} (1 - \textrm{PER}_i),
    \label{eq:24}
\end{equation}
\begin{equation}
    p_{\mathcal{I}} \triangleq \prod_{i \in \mathcal{I}} \left(1 - \textrm{PER}_i\right) \prod_{j \not\in \mathcal{I}} \textrm{PER}_j,
    \label{eq:25}
\end{equation}
\begin{equation}
    p_{K} \triangleq \sum_{k=K+1}^N \sum_{\mathcal{I} \in \mathcal{F}_k} p_{\mathcal{I}} = 1 - \left(p_{\mathcal{F}_{0}} +  \sum_{k = 1}^{K} \sum_{\mathcal{I} \in \mathcal{F}_k} p_{\mathcal{I}} \right).
    \label{eq:26}
\end{equation}

The upper bound in \eqref{eq:23} follows from the fact that $\varepsilon_{\mathcal{I}} \leq \varepsilon_{\mathcal{F}_N} = \textrm{tr}\left(\bm{C}_{\bm{\theta}} \right)$, where $\mathcal{F}_N = \emptyset$ represents the case of all errors. In addition, $K \in \{1,\ldots,N-1\}$ denotes the number of incorrect packets that are allowed in the decoding process before estimating $\bm{\theta}$. As an illustrative example, if $K = 2$, only two packets can contain errors; otherwise, the estimation is not carried out. Besides that, note that equality in \eqref{eq:23} is achieved for $K = N - 1$.

As a result, we can rewrite problem $(P1)$ as the minimization of the MSE upper bound $\bar{\varepsilon}_K$:
\begin{equation}    
    (P2): \{\bm{V}^{\star},\bm{n}^{\star}\} = \underset{\stackrel{\bm{V} \in \{0,1\}^{N \times M}}{ \bm{n} \in \mathbb{N}^M}}{\textrm{argmin}} \, \, \bar{\varepsilon}_K \quad \textrm{s.t.} \quad \bm{V} \bm{V}^{\textrm{T}} = \mathbf{I}_N, \quad \textrm{tr}\left(\bm{V}\bm{P}\bm{V}^{\textrm{T}}\right) \leq \delta.
    \label{eq:27}
\end{equation}

Nevertheless, considering the large number of devices involved in mMTC networks and the fact that, for practical systems, the individual error probabilities $\textrm{PER}_i$ tend to be small, the terms corresponding to $K > 1$ vanish quickly. This is because the second product in $p_{\mathcal{I}}$ tends to zero for small $\textrm{PER}_i$ and increasing $K$. Hence, working with small values of $K$ can yield a tight approximation. The accuracy of this bound will be further studied and properly justified through simulations in Section~\ref{sec:6.4}. Thus, although problem (P2) constitutes a worse-case scenario (minimization of an upper bound), its solution approaches that of (P1) even for a small $K$.

Finally, the next step is to find the solution of the problem in \eqref{eq:27}, i.e., the optimal values $\bm{V}^{\star}$ and $\bm{n}^{\star}$. Unfortunately, an analytic closed-form expression cannot be found, even for a small $N$ and $M$. That is why, in order to find the solution to the previous problem, we will make use of greedy iterative methods \cite{Cha20}. In that sense, first, we will consider the separate optimization of $\bm{V}$ and $\bm{n}$ and, later, we will concentrate on the joint optimization. Note that the latter will lead to better performance, yet at the cost of more computational complexity.

\subsection{Separate Optimization} \label{sec:4.1}
Problem (P2) can be decomposed into two optimizations, where $\bm{V}$ and $\bm{n}$ are optimized separately. On the one hand, given the number of bits $\bm{n}$, (P2) yields
\begin{equation}    
    (P2.1): \bm{V}^{\star} = \underset{\bm{V} \in \{0,1\}^{N \times M}}{\textrm{argmin}} \, \, \bar{\varepsilon}_K \quad \textrm{s.t.} \quad \bm{V}\bm{V}^{\textrm{T}} = \mathbf{I}_N, \quad \textrm{tr}\left(\bm{V}\bm{P}\bm{V}^{\textrm{T}}\right) \leq \delta,
    \label{eq:28}    
\end{equation}
which can be solved sequentially. At each step, we consider that $N - 1$ rows are given and, thus, the search concerns the remaining row \cite{Fan08a}. For instance, considering the first step, the matrix $\bm{V}$ can be expressed as
\begin{equation}
    \bm{V} = [\bm{v}_1^{\textrm{T}} \, \tilde{\bm{V}}_1^{\textrm{T}}]^{\textrm{T}},
    \label{eq:29}
\end{equation}
where $\bm{v}_1 \in \{0,1\}^{1 \times M}$ is the first row and $\tilde{\bm{V}}_1 = [\bm{v}_2^{\textrm{T}}, \ldots, \bm{v}_N^{\textrm{T}}]^{\textrm{T}} \in \{0,1\}^{(N-1) \times M}$ are the rest of fixed rows. Since each row can only contain one non-zero element and all rows must be different (constraints imposed by $\bm{V}\bm{V}^{\textrm{T}} = \mathbf{I}_N$), the optimal $\bm{v}_1$ is found by one-dimensional search:
\begin{equation}    
    (P2.1s): \bm{v}_1^{\star} = \underset{\bm{v}_1 \in \{0,1\}^{1 \times M}}{\textrm{argmin}} \, \, \bar{\varepsilon}_K
    \quad \textrm{s.t.} \quad [\bm{v}_1^{\textrm{T}} \, \tilde{\bm{V}}_1^{\textrm{T}}]^{\textrm{T}}[\bm{v}_1^{\textrm{T}} \, \tilde{\bm{V}}_1^{\textrm{T}}] = \mathbf{I}_N, \quad \textrm{tr}\left([\bm{v}_1^{\textrm{T}} \, \tilde{\bm{V}}_1^{\textrm{T}}]^{\textrm{T}}\bm{P}[\bm{v}_1^{\textrm{T}} \, \tilde{\bm{V}}_1^{\textrm{T}}]\right) \leq \delta.
    \label{eq:30}    
\end{equation}

This operation is then repeated for all rows in the same way as it has been done with the first row, and going back to the first one, until convergence is reached or when the MSE reduction is below a given threshold. Note that, as the optimization criterion is the minimization of the MSE, and one of the possibilities is to keep the same values (for $\bm{v}_i$) from the previous iteration, the new variable selection will always improve (i.e., decrease) the MSE or, in the worst case, keep the same. Therefore, convergence is always assured since we generate a monotonous decreasing sequence of values for the MSE.

In addition, as the search space is discrete, a closed-form solution cannot be derived. Consequently, the solution is found numerically by exhaustive search. In this case, the complexity grows linearly with the number of search elements since to solve \eqref{eq:30}, we only need to evaluate $M - N + 1$ different values (i.e., the number of silent sensors that can be potentially activated).

It is noteworthy that, since the MSE defined in \eqref{eq:20} is invariant to row permutations of the selection matrix $\bm{V}$ \cite{Kay93}, it can be shown that the total MSE $\varepsilon$ and the upper bound $\bar{\varepsilon}_K$ are also invariant to row re-ordering. For instance, if $M = 3$ and $N = 2$, the results with $\bm{v}_1 = [0 1 0]$ and $ \bm{v}_2 = [1 0 0]$ are equivalent to those with $\bm{v}_1 = [1 0 0]$ and $\bm{v}_2 = [0 1 0]$. 

Likewise, when the selection matrix $\bm{V}$ is known, problem (P2) reads as
\begin{equation}
    (P2.2): \quad \bm{n}^{\star} = \underset{\bm{n} \in \mathbb{N}^M}{\textrm{argmin}} \, \, \bar{\varepsilon}_K,
    \label{eq:31}
\end{equation}
which can also be solved through a sequential (and iterative) procedure. Thereby, considering that the last $M -1$ elements in $\bm{n}$ are given, at each step the optimization yields
\begin{equation}
    (P2.2s): \quad n_1^{\star} = \underset{n_1 \in \mathbb{N}}{\textrm{argmin}} \, \, \bar{\varepsilon}_K,
    \label{eq:32}
\end{equation}
procedure that is also repeated for all elements in $\bm{n}$ and iterated until converge is reached. As before, convergence is guaranteed since each iteration yields the same or a lower MSE (i.e., we can always keep the previous value for $n_i$, which will maintain the MSE, or select a new one that decreases the estimation error). In that sense, similarly to (P2.1s), we employ exhaustive search to solve \eqref{eq:32} given that here the search space is also discrete and a closed-form solution cannot be found. Thereby, since the number of quantization bits is usually limited (i.e., $n_i \leq B$, where $B$ is the maximum number of quantization bits), the number of possibilities we evaluate is given by this upper bound $B$.

Finally, the two separate optimizations are sequentially alternated in order to find a (sub-optimal) stationary solution. This way, the resulting MSE will monotonically decrease until a local minimum is attained (or when the MSE is below a given threshold). The entire procedure is described in Algorithm 1.

\begin{algorithm}[t]
\caption{Separate (alternate) optimization to solve $(P2)$}          
\label{alg:1}                         
\begin{algorithmic}[1]
\State Initialize\footnotemark $\bm{V}$ and $\bm{n}$
	\Repeat
	    \Repeat
	    \For{$i = 1:N$}
		    \State Solve $(P2.1s)$ through one-dimensional search to find $\bm{v}_i^{\star}$
		    \State $\bm{v}_i \gets \bm{v}_i^{\star}$
	    \EndFor
		\Until{Convergence is reached (or the MSE is below a given threshold)}
		\Repeat
		\For{$j = 1:M$}
		    \State Solve $(P2.2s)$ through one-dimensional search to find $n_j^{\star}$
		    \State $n_j \gets n_j^{\star}$
	    \EndFor
		\Until{Convergence is reached (or the MSE is below a given threshold)}
	\Until{Convergence is reached (or the MSE is below a given threshold)}
\end{algorithmic}
\end{algorithm}

\subsection{Joint Optimization} \label{sec:4.2}
Instead of considering the separate approach described above, we can search for the solution of $\bm{V}$ and $\bm{n}$ simultaneously through a similar iterative procedure. More specifically, we will consider that $N-1$ rows of $\bm{V}$ and $M-1$ elements of $\bm{n}$ are given. Without loss of generality, we discuss the optimization of the first row of $\bm{V}$ like in \eqref{eq:29}, and the corresponding element in $\bm{n}$, which depends on the device selection. Hence, at each step, we will look for the optimal solution through a two-dimensional search:
\begin{equation}
\begin{aligned}
     & (P2s): &\{\bm{v}_1^{\star},n_{i(\bm{v}_1)}^{\star}\} = &\underset{\stackrel{\bm{v}_1 \in \{0,1\}^{1 \times M}}{ n_{i(\bm{v}_1)} \in \mathbb{N}}}{\textrm{argmin}} \, \, \bar{\varepsilon}_K, \\ 
    && \textrm{s.t.} \quad &[\bm{v}_1^{\textrm{T}} \, \tilde{\bm{V}}_1^{\textrm{T}}]^{\textrm{T}}[\bm{v}_1^{\textrm{T}} \, \tilde{\bm{V}}_1^{\textrm{T}}] = \mathbf{I}_N, \quad \textrm{tr}\left([\bm{v}_1^{\textrm{T}} \, \tilde{\bm{V}}_1^{\textrm{T}}]^{\textrm{T}}\bm{P}[\bm{v}_1^{\textrm{T}} \, \tilde{\bm{V}}_1^{\textrm{T}}]\right) \leq \delta,
    \label{eq:33}
\end{aligned}
\end{equation}
where the index $i(\bm{v}_1)$ represents the sensor selected in $\bm{v}_1$. For instance, if $M = 5$ and $\bm{v}_1 = [0 1 0 0 0]$, then $i(\bm{v}_1) = 2$. Therefore, since we only optimize the quantization bits of the selected devices, the dimensionality of that search reduces from $M$ to $N$ (the same applies in the case of the separate optimization). The rest of $M - N$ quantization bits can remain unspecified. As before, the other rows of $\bm{V}$ and elements of $\bm{n}$ can be found through the same procedure (i.e., two-dimensional search), which is sequentially repeated until the MSE converges.

\footnotetext{One possibility is to initialize the selection matrix $\bm{V}$ randomly and set the vector of quantization bits $\bm{n}$ to $n_{\textrm{min}} \mathbf{1}_M$, where $n_{\textrm{min}}$ is the minimum number of quantization bits. Note that, although this will be the case we consider in this work, our algorithm is independent of the variable initialization.}

Compared to the separate case, a higher computational complexity is required. In particular, for a maximum number of bits $B \geq n_i$, each iteration in the separate optimization requires $N(M-N + 1)$ trials for the selection matrix and $B  N$ trials for the quantization. Contrarily, $N (M - N + 1) B N$ trials are needed in the joint case. To take into account the total number of iterations, in the simulations section, we will show the execution time required by each approach.

\section{Extension to Temporal Correlation} \label{sec:5}
As mentioned before, the previous analysis can be extended to the case of scenarios where the parameters $\bm{\theta}$ to be estimated vary over time with a given temporal correlation (in addition to the already mentioned spatial correlation). The temporal evolution of the parameters $\bm{\theta}$ will be modeled through first-order Markov processes \cite{Mse08,Fre19}. 

In this situation, the estimation strategy to be designed at the CN will have memory to exploit that temporal correlation. In particular, given the MSE criterion, we make use of Kalman filters to estimate the parameter vector $\bm{\theta}$, which is optimum under the Gaussian assumption.

\subsection{System Model} \label{sec:5.1}
We start by considering an observation time of $T$ frames. Thus, now the measurements $x_i$ depend on the frame $t \in \{1,\ldots,T\}$, i.e., $x_i(t) \in \mathbb{R}$ is the measurement of the $i$-th sensor at the $t$-th frame. Following the model in \eqref{eq:1}, at frame $t$ we have:
\begin{equation}
    \bm{x}(t) = \bm{\theta}(t) + \bm{w}(t) \in \mathbb{R}^M, 
    \label{eq:34}
\end{equation}
where $\bm{x}(t) = [x_1(t), \ldots, x_M(t)]^{\textrm{T}}$ is the set of measurements, $\bm{\theta}(t) \in \mathbb{R}^M$ is the parameter vector, and $\bm{w}(t) \in \mathbb{R}^M$ is the measurement noise vector. Like before, we assume $\bm{\theta}(t)$ to be Gaussian distributed with (known) mean $\bm{\mu}_{\bm{\theta}} (t)$ and covariance matrix $\bm{C}_{\bm{\theta}} (t)$, i.e., $\bm{\theta}(t) \sim \mathcal{N}(\bm{\mu}_{\bm{\theta}}(t),\bm{C}_{\bm{\theta}}(t))$. Besides, we assume that $\bm{w}(t)$ is independent of $\bm{\theta}(t)$, temporally uncorrelated, and Gaussian distributed with zero mean and (known) covariance matrix $\bm{C}_{\bm{w}}(t)$, i.e., $\bm{w}(t) \sim \mathcal{N}(\bm{0},\bm{C}_{\bm{w}}(t))$.

On the other hand, the temporal evolution of $\bm{\theta}(t)$ can be expressed with the following first-order Markov model \cite{Kay93}:
\begin{equation}
    \bm{\theta}(t) = \bm{F}(t) \bm{\theta}(t - 1) + \bm{\nu}(t),
    \label{eq:35}
\end{equation}
where $\bm{F}(t) \in \mathbb{R}^{M \times M}$ is the transition matrix, which is assumed to be known, and $\bm{\nu}(t) = [\nu_1(t),\ldots,\nu_M(t)]^{\textrm{T}} \in \mathbb{R}^M$ is the process noise, uncorrelated and independent of $\bm{w}(t)$ and $\bm{\theta}(t)$. We also consider that $\bm{\nu}(t)$ follows a Gaussian distribution with zero mean and (known) covariance matrix $\bm{C}_{\bm{\nu}}(t)$, i.e., $\bm{\nu}(t) \sim \mathcal{N}(\bm{0},\bm{C}_{\bm{\nu}}(t))$.

Thereby, now the quantized observations also depend on the frame index, i.e., $y_i(t) \in \mathbb{R}$ is the quantized version of $x_i(t)$ at the $t$-th frame. Hence, the set of quantized observations $\bm{y}(t) = [y_1(t),\ldots,y_M(t)]^{\textrm{T}} \in \mathbb{R}^M$ reads as (cf. \eqref{eq:4})
\begin{equation}
\bm{y}(t) = \bm{x}(t) + \bm{\eta}(t) \in \mathbb{R}^M,
\label{eq:36}
\end{equation}
where $\bm{\eta}(t) = [\eta_1(t),\ldots,\eta_M(t)]^{\textrm{T}} \in \mathbb{R}^M$ is the quantization noise vector. Note that the elements $\eta_i(t)$ are uniformly distributed within the quantization interval $\Delta_i(t)= S_i(t)/2^{n_i(t)}$, i.e., $\eta_i \sim \mathcal{U}(-\Delta_i(t)/2,\Delta_i(t)/2)$. Accordingly, $n_i(t)$ is now the number of quantization bits, which may vary in time according to the MSE minimization.

This way, considering errors in the communication, the set of available quantized observations at the CN can be written as (cf. \eqref{eq:18})
\begin{equation}
    \bm{z}_{\mathcal{I}(t)}(t) = \bm{A}_{\mathcal{I}(t)}\bm{V}(t)\bm{y}(t) \in \mathbb{R}^{\vert\mathcal{I}(t)\vert},
    \label{eq:37}
\end{equation}
where the selection matrix $\bm{V}(t) \in \{0,1\}^{N \times M}$ is also allowed to change over the different frames to optimize the MSE. In addition, note that $\mathcal{I}(t)$ is the set of active sensors with correctly decoded messages at the $t$-th frame and $\bm{A}_{\mathcal{I}(t)}$ is the corresponding matrix indicating the absence of errors. 

\subsection{Parameter Estimation} \label{sec:5.2}
Since we consider the MSE as the design criterion, a suitable choice for the MMSE estimator is the linear Kalman filter, which is optimum under the Gaussian assumption \cite{Mse12,She14,Das15}. Thereby, now the estimation will consist in two-steps, namely prediction and correction. As discussed in \cite{Kay93}, the MMSE estimate of $\bm{\theta}(t)$ at the $t$-th frame, assuming that the sensors with correctly detected signals are indexed by $\mathcal{I}(t)$, can be obtained recursively:
\begin{equation}    
    \hat{\bm{\theta}}_{\mathcal{I}(t)}(t \vert t) = \mathbb{E}[\bm{\theta}(t) \vert \bar{\bm{z}}(1),\ldots,\bar{\bm{z}}(t - 1), \bm{z}_{\mathcal{I}(t)}(t)] = \underbrace{\hat{\bm{\theta}}(t \vert t - 1)}_{\textrm{prediction}} + \underbrace{\mathbb{E}[\bm{\theta}(t) \vert \tilde{\bm{z}}_{\mathcal{I}(t)}(t)]}_{\textrm{correction}},
    \label{eq:38}    
\end{equation}
where $\bar{\bm{z}}(1)\in \mathbb{R}^{C(1)},\ldots,\bar{\bm{z}}(t-1)\in \mathbb{R}^{C(t-1)}$ are the sets of $C(1),\ldots,C(t-1)$ available observations (i.e., selected and correctly decoded) at previous frames $1,\ldots,t-1$; $\hat{\bm{\theta}}(t \vert t -1) \in \mathbb{R}^{\vert \mathcal{I}(t) \vert}$ is the prediction of $\bm{\theta}(t)$ given the available observations until frame $t-1$; and $\tilde{\bm{z}}_{\mathcal{I}(t)}(t) = \bm{z}_{\mathcal{I}(t)}(t) - \hat{\bm{z}}_{\mathcal{I}(t)}(t \vert t - 1) \in \mathbb{R}^{\vert \mathcal{I}(t) \vert}$ is the innovation in $\bm{z}_{\mathcal{I}(t)}(t)$ (i.e., information provided by $\bm{z}_{\mathcal{I}(t)}(t)$ but not by the past measurements $[\bar{\bm{z}}(1),\ldots,\bar{\bm{z}}(t - 1)]$). Besides, $\hat{\bm{z}}_{\mathcal{I}(t)}(t \vert t - 1) \in \mathbb{R}^{\vert \mathcal{I}(t) \vert}$ represents the prediction of $\bm{z}_{\mathcal{I}(t)}(t)$ given the observations until frame $t-1$.

Thereby, we first wish to predict the parameters at the $t$-th frame based on the observations from the previous frames, namely the first term in \eqref{eq:38}. In particular, since the innovation sequences contain the same information as the original data, it can be shown that the MMSE prediction of $\bm{\theta}(t)$ reads as \cite{Kay93}:
\begin{equation}
    \hat{\bm{\theta}}(t\vert t - 1) = \mathbb{E}[\bm{\theta}(t) \vert \tilde{\bm{z}}(1),\ldots,\tilde{\bm{z}}(t - 1)] = \bm{F}(t)\hat{\bm{\theta}}(t-1 \vert t - 1),
    \label{eq:39}
\end{equation}
where $\tilde{\bm{z}}(1) = \bar{\bm{z}}(1) - \hat{\bm{z}}(1 \vert 0) \in \mathbb{R}^{C(1)},\ldots, \tilde{\bm{z}}(t-1) = \bar{\bm{z}}(t-1) - \hat{\bm{z}}(t-1 \vert t-2) \in \mathbb{R}^{C(t-1)}$ are the sets of $C(1),\ldots,C(t-1)$ available innovations at previous frames $1,\ldots,t-1$.

Next, we need to correct the prediction with the current observation, i.e., the second term in \eqref{eq:38}. Following the derivations in \cite{Kay93}, we have
\begin{equation}
    \mathbb{E}[\bm{\theta}(t) \vert \tilde{\bm{z}}_{\mathcal{I}(t)}(t)] = \bm{K}_{\mathcal{I}(t)}(t) \left(\bm{z}_{\mathcal{I}(t)}(t) - \hat{\bm{z}}_{\mathcal{I}(t)}(t \vert t - 1)\right),
    \label{eq:40}
\end{equation}
where $\hat{\bm{z}}_{\mathcal{I}(t)}(t \vert t - 1) = \bm{A}_{\mathcal{I}(t)}\bm{V}(t) \hat{\bm{\theta}}(t\vert t - 1)$ is the MMSE prediction of $\bm{z}_{\mathcal{I}(t)}(t)$ and $\bm{K}_{\mathcal{I}(t)}(t) \in \mathbb{R}^{M \times \vert \mathcal{I}(t) \vert}$ is the so-called Kalman gain and is given by \cite{Kay93}
\begin{equation}
    \bm{K}_{{\mathcal{I}(t)}}(t) = \bm{\Sigma}(t \vert t - 1) \bm{V}(t)^{\textrm{T}} \bm{A}_{\mathcal{I}(t)}^{\textrm{T}} \Big(\bm{A}_{\mathcal{I}(t)}\bm{V}(t)\big(\bm{\Sigma}(t \vert t - 1) + \bm{C}_{\bm{w}}(t) + \bm{C}_{\Delta}(t) \big)\bm{V}(t)^{\textrm{T}}\bm{A}_{\mathcal{I}(t)}^{\textrm{T}}\Big)^{-1},
    \label{eq:41}
\end{equation}
with $\bm{C}_{\Delta}(t) = \textrm{diag}\left(\Delta_1^2(t)/12,\ldots,\Delta_M^2(t)/12\right)$. Note that $\bm{\Sigma}(t \vert t - 1)\in \mathbb{R}^{M \times M}$ is the error covariance matrix of the parameter prediction:
\begin{equation}
    \bm{\Sigma}(t \vert t - 1) = \bm{F}(t) \bm{\Sigma}(t - 1 \vert t - 1)\bm{F}^{\textrm{T}}(t) + \bm{C}_{\bm{\nu}}(t),
    \label{eq:42}
\end{equation}
with $\bm{\Sigma}(t - 1 \vert t - 1) \in \mathbb{R}^{M \times M}$ being the covariance matrix of the parameter estimate at frame $t-1$. In general, at the $t$-th frame, the following recursive expression holds:
\begin{equation}
    \bm{\Sigma}_{\mathcal{I}(t)}(t \vert t) = \left(\mathbf{I}_M - \bm{K}_{{\mathcal{I}(t)}}(t)\bm{A}_{\mathcal{I}(t)} \bm{V}(t)\right) \bm{\Sigma}(t \vert t - 1).
    \label{eq:43}
\end{equation}

As a result, the MSE yields (cf. \eqref{eq:21})
\begin{equation}    
    \varepsilon(t) \triangleq \sum_{k = 0}^{N}\sum_{\mathcal{I}(t) \in \mathcal{F}_k}  \varepsilon_{\mathcal{I}(t)}(t) \prod_{i \in \mathcal{I}(t)} \left(1 - \textrm{PER}_i(t)\right) \prod_{j \not\in \mathcal{I}(t)}  \textrm{PER}_j(t),
    \label{eq:44}    
\end{equation}
where $\textrm{PER}_i(t)$ is the packet error rate at the $t$-th frame, which is given in \eqref{eq:9} when considering $n_i(t)$ information bits. Likewise, following expression \eqref{eq:20}, $\varepsilon_{\mathcal{I}(t)}(t)$ can be obtained by computing the trace of the matrix in \eqref{eq:43}.

Finally, since the goal here is to minimize the MSE in \eqref{eq:44}, we can simply follow the procedures described in Section~\ref{sec:4} to obtain a sub-optimal solution for $\bm{V}(t)$ and $\bm{n}(t) = [n_1(t),\ldots,n_M(t)]^{\textrm{T}}$. In short, the MSE in \eqref{eq:44} can be upper bounded by $\bar{\varepsilon}_K(t)$ (cf. \eqref{eq:23}) and the resulting problem can be solved using a separate or a joint optimization procedure. Besides, note that the problem constraints could be formulated for a variable number of slots $N(t)$, i.e., $\bm{V}(t)\bm{V}(t)^{\textrm{T}} = \mathbf{I}_{N(t)}$, and a variable matrix transmit power $\bm{P}(t)$. However, this analysis is beyond the scope of this work, and we will only consider the case of $N(t) = N \, \forall t$ and $\bm{P}(t) = \bm{P} \, \forall t$.

\section{Numerical Simulations} \label{sec:6}
In this section, several simulations are presented to illustrate the performance of our approach. More specifically, the resulting MSE $\varepsilon$ from \eqref{eq:21} after the device selection and information quantization is analyzed for different setups. Later, we will concentrate on the results and the evolution of the MSE $\varepsilon(t)$ defined in \eqref{eq:44} for the case of temporally correlated data. 

To that end, we will consider a realistic mMTC network for our study. In particular, we will use the parameters and guidelines specified by the 3GPP and ITU standards \cite{3GPP38214,ITU09}. In addition, we will distinguish between the case where the measurements are generated synthetically and the case where they are obtained from the database collected by the Intel Berkeley Research Lab \cite{Bod04}. That is why, in the following, we dedicate initial subsections to discuss all these practical issues and to describe the different datasets. 

\subsection{Practical Issues} \label{sec:6.1}
Throughout all simulations, we consider the micro-urban scenario described in \cite{ITU09} with $P_i = P = 0$ dBm and $N_o = -174$ dBm/Hz. Accordingly, the power matrix results $\bm{P} = P \mathbf{I}_M$ and, therefore, the power consumption and power reduction are given by $NP$ and $(M - N)P$, respectively. This way, the ratio of power consumption (i.e., ratio between consumed power and total power) and reduction (i.e., ratio between reduced power and total power) are given by the ratio of active sensors $N/M$ and silent sensors $(M - N)/M$, respectively. In addition, for the optimization problem to be feasible, we set the threshold $\delta$ to $NP$\footnote{ In this particular case, the effect of the power constraint disappears since $\textrm{tr}(\bm{V}\bm{P}\bm{V}^{\textrm{T}})=P \cdot \textrm{tr}(\bm{V} \mathbf{I}_M \bm{V}^{\textrm{T}}) = NP$.}.

Besides, since here we will consider that the information to be transmitted is rather small (e.g., temperature), sensors will employ a single resource element, i.e., $T_s \approx 71.4$ $\mu\textrm{s}$ \cite{3GPP38214} and $B_s = 15 $ kHz \cite{Nok16}. The channel gains $G_i$ are computed for a power law path-loss, i.e., 
\begin{equation}
    G_i = D_i^{-\alpha} \sigma_{g}^2,
    \label{eq:45}
\end{equation}
where $D_i$ is the distance between sensor $i$ and the CN, $\alpha = 3$ is the decay exponent, and $\sigma_g^2 = 1$ is the power of the fading coefficient. Note that this model is defined for single-antenna devices but it could be extended to the multi-antenna case by introducing the corresponding diversity gain in expression \eqref{eq:45}. For more information, please refer to \cite{Van04}.

Regarding the spatial distribution, we consider sensors to be uniformly distributed around the CN within a disk of radius $50$ m. This small area helps us to reduce the computational complexity of the numerical methods, as we can use smaller values of $M$ while preserving the high spatial density of sensors and, thus, the high spatial correlation between the sensed data. Note that this is also the case of the Intel dataset, where the exact positions of the sensors are not randomly generated but extracted from the actual deployment in the laboratory \cite{Bod04}.

\subsection{Synthetic Data} \label{sec:6.2}
As an example of application, we consider temperature as the phenomenon to be measured. Note that, although in the upcoming simulations we focus on a single phenomenon (temperature), the parameters $\theta_i$ measured by the sensors are the different realizations of this phenomenon at the different locations of the sensors. In fact, our formulation would still be valid for the case of multiple phenomena (e.g., temperature, humidity, etc.) if each sensor only takes a single scalar measurement of just one of them. In that case, the presented technique could exploit the potential correlation between these phenomena.

In the following, we assume a measurement noise with $\bm{C}_{\bm{w}} = \textrm{diag}(\bm{\sigma}_w^2)$, $\bm{\sigma}_w = [\sigma_{w,1}^2 , \ldots, \sigma_{w,M}^2]$, and $ \sigma_{w,i} \in [0, 10] \SI{}{\celsius}$. Also, we make use of an exponential model for the covariance matrix $\bm{C}_{\bm{\theta}}$ \cite{Vur04} with $\sigma_i = \sigma = \SI{10}{\celsius}$ and
\begin{equation}
    K(d_{i,j}) = \textrm{exp}(-d_{i,j}/\phi),
    \label{eq:46}
\end{equation}
where $\phi > 0$ controls the degree of spatial correlation and $d_{i,j}$ is the distance between sensors $i$ and $j$. For the sake of simplicity, we define $\varphi = \textrm{exp} (-1/\phi) \in [0,1]$ to study the impact of this degree, i.e., $\varphi = 0$ is the case of no correlation and $\varphi = 1$ is the case of identical parameters.

In the case of a system with memory, we model the transition matrix as $\bm{F}(t) = \psi \mathbf{I}_M \, \forall t$, where $0 \leq \psi \leq 1$ represents the coefficient measuring the temporal correlation between consecutive observations. Accordingly, we consider that the covariance matrix of the process noise is given by $\bm{C}_{\bm{\nu}}(t) = (1 - \psi^2) \bm{C}_{\bm{\theta}} (t) \, \forall t$ \cite{Fre19}. Likewise, the impact of $\psi$ on the MSE will be also analyzed. In all cases, we will consider a period of time of $T = 50$ frames.

Finally, recall that the number of quantization bits $\bm{n}$ actually depends on the MCS available for each device, i.e., $R_i$ and $L_i$ (cf. \eqref{eq:8}). In particular, the set of possible values for $\bm{n}$ can be derived from Table 5.2.2.1 in \cite{3GPP38214}, where the maximum number of (information) bits is limited to $8$. Accordingly, the corresponding functions $\mathcal{C}_i(\cdot)$ from \eqref{eq:9} are obtained following the indications of the LTE standard \cite{Gho11}.

\subsection{Real Data} \label{sec:6.3}
To further evaluate the performance of our approach, in this paper we also consider the set of measurements obtained by the Intel Berkeley Research Lab \cite{Bod04}. This data base consists of $M = 54$ deployed sensors transmitting their data (temperature) to the CN every $31$ seconds (s) during $D = 38$ days. As a result, there are $2.3$ million readings available. However, since the database is not complete (not all time slots contain samples), we consider that days are divided into intervals of duration $H = 900$ s, so that each time instant contains at least one measurement.

Note that each of these measurements corresponds to the set of $Q$ observations collected at day $d \in \{1,\ldots,D\}$ and time instant $t \in \{1,\ldots, I/H \}$, i.e., 
\begin{equation}
    \bm{X}(d,t)= [\bm{x}_1(d,t),\ldots,\bm{x}_Q(d,t)] \in \mathbb{R}^{M \times Q},
    \label{eq:47}
\end{equation}
where $I$ corresponds to the day duration (i.e., $86400$ s), and $\bm{x}_j(d,t) \in \mathbb{R}^M$ is the $j$-th vector of samples, with $j \in \{1,\ldots,Q\}$, collected at day $d$ during time instant $t$. Accordingly, each $\bm{x}_j(d,t)$ can be expressed with the following observation model (cf. \eqref{eq:34}):
\begin{equation}
    \bm{x}_j(d,t) = \bm{\theta}_j(d,t) + \bm{w}_j(d,t),
    \label{eq:48}
\end{equation}
where $\bm{\theta}_j(d,t) = \bm{\theta}(d,t) \in \mathbb{R}^M$ and $\bm{w}_j(d,t) \in \mathbb{R}^M$ are the $j$-th parameter vector and noise vector at day $d$ and time instant $t$. Similarly to before, we assume both vectors have Gaussian distributions: $\bm{\theta}(d,t) \sim \mathcal{N}\left(\bm{\mu}_{\bm{\theta}}(t), \bm{C}_{\bm{\theta}}(t)\right) $ $\forall j,d$ and $\bm{w}_j(d,t) \sim \mathcal{N}\left(\bm{0}, \bm{C}_{\bm{w}}(t)\right)$ $\forall j,d$. Note that we consider stationarity over the different days. 

Thereby, considering the temperature remains constant over an interval of $L = 2700$ s, the parameter $\bm{\theta}(d,t)$ can be obtained by averaging $\bm{X}(d,t)$ with a sliding window of size $J = L/H$ intervals centered at $t$:
\begin{equation}
    \bm{\theta}(d,t) = \frac{1}{JQ} \sum_{s = t - \frac{J - 1}{2}}^{t + \frac{J-1}{2}} \sum_{j = 1}^Q \bm{x}_j(d,s).
    \label{eq:49}
\end{equation}

Accordingly, the observation noise $\bm{w}_j(d,t)$ can be obtained by subtracting $\bm{\theta}(d,t)$ to each of the observations in \eqref{eq:47}: 
\begin{equation}
    \bm{w}_j(d,t) = \bm{x}_j(d,t) - \bm{\theta}(d,t).
    \label{eq:50}
\end{equation}

This way, the statistical moments of $\bm{\theta}(d,t)$ and $\bm{w}_j(d,t)$ can be obtained as follows:
\begin{equation}
    \bm{\mu}_{\bm{\theta}} (t) = \frac{1}{D} \sum_{d = 1}^{D} \bm{\theta}(d,t),
    \label{eq:51}
\end{equation}
\begin{equation}
    \bm{C}_{\bm{\theta}} (t) = \frac{1}{D} \sum_{d = 1}^{D} (\bm{\theta}(d,t) - \bm{\mu}_{\bm{\theta}} (t)) (\bm{\theta}(d,t) - \bm{\mu}_{\bm{\theta}} (t))^{\textrm{T}},
    \label{eq:52}
\end{equation}
\begin{equation}
    \bm{C}_{\bm{w}} (t) = \frac{1}{D Q} \sum_{d = 1}^{D} \sum_{j = 1}^Q \bm{w}_j(d,t) \bm{w}_j(d,t)^{\textrm{T}}.
    \label{eq:53}
\end{equation}

On the other hand, regarding the dynamical model, we have (cf. \eqref{eq:35}):
\begin{equation}
    \bm{\theta}(d,t) = \bm{F}(t) \bm{\theta}(d,t-1) + \bm{\nu}(d,t),
    \label{eq:54}
\end{equation}
where $\bm{\nu}(d,t) \in \mathbb{R}^M$ is the process noise at day $d$ and time instant $t$. We also assume that $\bm{\nu}(d,t) \sim \mathcal{N}\left(\bm{0}, \bm{C}_{\bm{\nu}}(t)\right)$. In addition, we assume that the transition matrix is a scaled identity, i.e., $\bm{F}(t) = \alpha (t) \mathbf{I}_M$. Thereby, the factor $\alpha(t)$ can be obtained as follows:
\begin{equation}
    \alpha(t) = \frac{1}{D}\sum_{d = 1}^{D}\frac{\bm{1}_M^{\textrm{T}}\bm{\theta}(d,t)}{\bm{1}_M^{\textrm{T}}\bm{\theta}(d,t - 1)}.
    \label{eq:55}
\end{equation}

Finally, the process noise is given by
\begin{equation}
    \bm{\nu}(d,t) = \bm{\theta}(d,t) - \alpha(t) \bm{\theta}(d,t - 1),
    \label{eq:56}
\end{equation}
with covariance matrix
\begin{equation}
    \bm{C}_{\bm{\nu}}(t) = \frac{1}{D} \sum_{d = 1}^{D} \bm{\nu}(d,t) \bm{\nu}(d,t)^{\textrm{T}}.
    \label{eq:57}
\end{equation}

\subsection{Sensor Selection and Distributed Quantization with Synthetic Data} \label{sec:6.4}
\begin{figure}[t]
    \centering 
    \includegraphics[scale = 1]{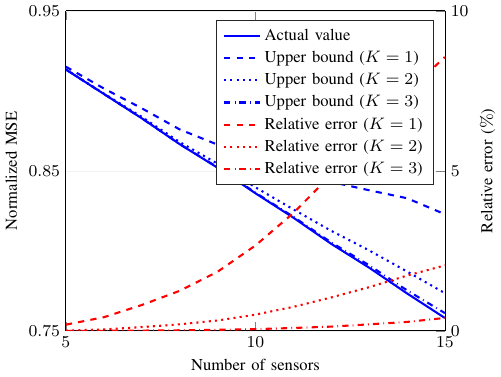} 
    \caption{Original normalized MSE $\varepsilon / \textrm{tr}(\bm{C}_{\bm{\theta}})$ and normalized upper bound $\bar{\varepsilon}_K / \textrm{tr}(\bm{C}_{\bm{\theta}})$ (left) and relative error (right) versus number of sensors $M$.}
    \label{fig:3}
\end{figure}

We first start this subsection by evaluating the accuracy of the upper bound $\bar{\varepsilon}_K$ defined in \eqref{eq:23}. For this task, in Fig.~\ref{fig:3}, we present this magnitude with respect to (w.r.t.) $M$ and different values of $K$, together with the actual MSE $\varepsilon$ and the corresponding relative error. Recall that $K$ represents the number of incorrect packets we allow when computing the upper bound. 

For better visualization, both metrics are normalized by $\textrm{tr}\left(\bm{C}_{\bm{\theta}}\right)$, i.e., the MSE is bounded between $0$ and $1$. As mentioned before, small values of $K$, e.g., $K = 3$, provide a tight bound. Therefore, for a safe and consistent analysis, from now on we will consider $\bar{\varepsilon}_5$ as a suitable approximation for substituting $\varepsilon$ in the optimization defined in \eqref{eq:27}. The same reasoning holds for the optimization of $\varepsilon(t)$, i.e., $K = 5$ is used when solving the corresponding problem. 

\begin{figure*}[t]
\begin{multicols}{2}
    \centering 
    \includegraphics[scale = 1]{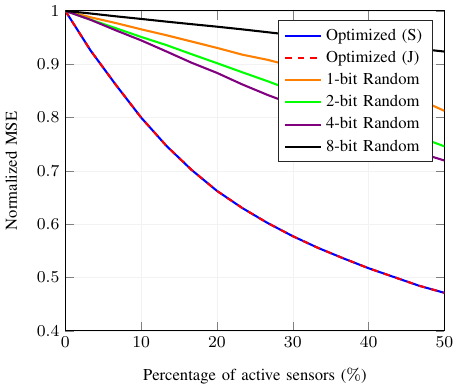} 
    \caption{Normalized MSE $\varepsilon / \textrm{tr}(\bm{C}_{\bm{\theta}})$ versus percentage of active sensors $N/M \cdot 100$ with $\varphi = 0.1$.}
    \label{fig:4}
    \includegraphics[scale = 1]{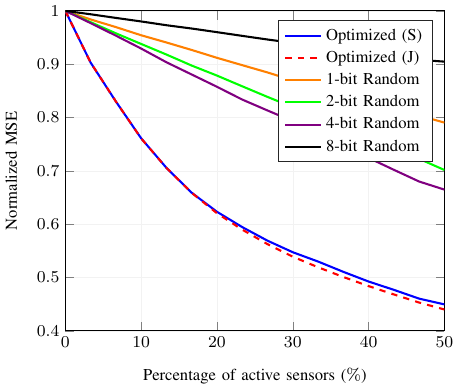} 
    \caption{Normalized MSE $\varepsilon / \textrm{tr}(\bm{C}_{\bm{\theta}})$ versus percentage of active sensors $N/M \cdot 100$ with $\varphi = 0.9$.}
    \label{fig:5}
\end{multicols}
\begin{multicols}{2}
    \centering
    \includegraphics[scale = 1]{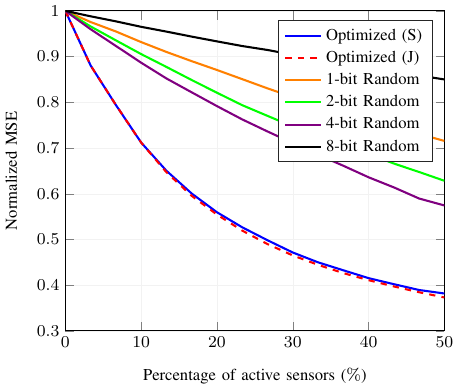} 
    \caption{Normalized MSE $\varepsilon / \textrm{tr}(\bm{C}_{\bm{\theta}})$ versus percentage of active sensors $N/M \cdot 100$ with $\varphi = 0.95$.}
    \label{fig:6}
    \includegraphics[scale = 1]{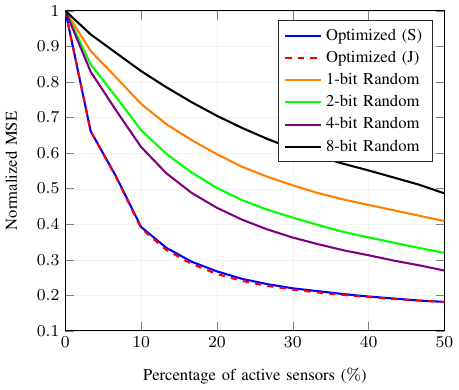}
    \caption{Normalized MSE $\varepsilon / \textrm{tr}(\bm{C}_{\bm{\theta}})$ versus percentage of active sensors $N/M \cdot 100$ with $\varphi = 0.99$.}
    \label{fig:7}
\end{multicols}
\end{figure*}

On the other hand, to study the performance of our estimation approach in the memoryless case, we will show the normalized MSE (NMSE), i.e., $\varepsilon / \textrm{tr}\left(\bm{C}_{\bm{\theta}}\right)$,  w.r.t. the percentage of active sensors, i.e., $N/M \cdot 100$, and different values of $\varphi$. The results are depicted in Fig.~\ref{fig:4}, Fig.~\ref{fig:5}, Fig.~\ref{fig:6}, and Fig.~\ref{fig:7} for $\varphi = 0.1$ (low correlation), $\varphi = 0.9$, $\varphi = 0.95$ (high correlation), and $\varphi = 0.99$ (almost identical observations), respectively. In all plots, we consider $M = 30$ sensors. The separate and joint optimizations are denoted by (S) and (J), respectively. Also, recall that $\bar{\varepsilon}_K$ is only used in the optimization step, while $\varepsilon$ is the value we actually show in all figures.

It can be seen that, in both cases (S and J), a better performance is obtained when the percentage of active sensors increases. This is not surprising as more sensors are allowed to transmit and, thus, more data can be retrieved. For instance, with $\varphi = 0.9$, the normalized MSE decreases from $0.76$ to $0.48$ for $10\%$ and $40 \%$ of active sensors (or consumed power), respectively. The same effect can be observed when the coefficient $\varphi$ that measures correlation increases, i.e., a smaller loss is attained for the same ratio $N/M$. This is because most of the relevant information is contained in less sensors. For instance, at $N/M = 0.1$ (i.e., $10 \%$ of power consumption), the MSE with $\varphi = 0.1$ and $\varphi = 0.99$ results in $0.79$ and $0.38$, respectively. 

Furthermore, to compare the performance of our system, the MSE of a random selection with a fixed number of quantization bits is also illustrated in Fig.~\ref{fig:4}, Fig.~\ref{fig:5}, Fig.~\ref{fig:6}, and Fig.~\ref{fig:7}. As expected, a higher error is obtained with this naive approach, especially for a large percentage of active sensors. Besides, note that, when increasing a lot the number of quantization bits (e.g., $8$ bits), the MSE also increases. The reason is that, given the presence of communication errors, a higher precision (i.e., more quantization bits) implies a weaker codeword protection (cf. \eqref{eq:8}). This translates into a poorer decoding and, therefore, into larger BER and PER. A larger MSE is also achieved when the number of bits is too low (e.g., $1$ bit), since the information sent by the sensors is not precise enough for a good estimation. Like before, when the correlation increases, the random approach yields less error since more sensors contain the same information.

\begin{figure*}[t]
\begin{multicols}{2}
    \centering 
    \includegraphics[scale = 1]{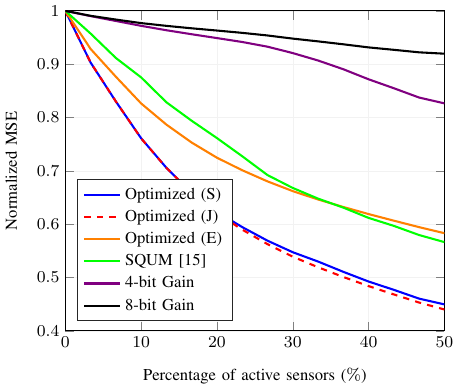} 
    \caption{Normalized MSE $\varepsilon / \textrm{tr}(\bm{C}_{\bm{\theta}})$ versus percentage of active sensors $N/M \cdot 100$ with $\varphi = 0.9$.}
    \label{fig:8}
    \includegraphics[scale = 1]{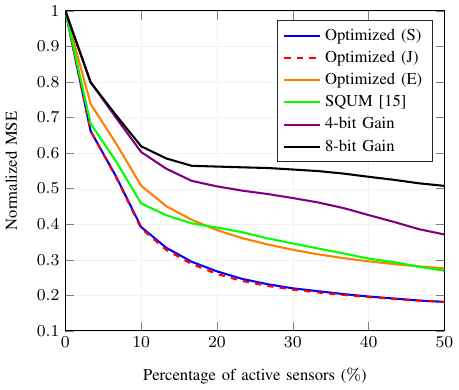} 
    \caption{Normalized MSE $\varepsilon / \textrm{tr}(\bm{C}_{\bm{\theta}})$ versus percentage of active sensors $N/M \cdot 100$ with $\varphi = 0.99$.}
    \label{fig:9}
\end{multicols}
\begin{multicols}{2}
    \centering 
    \includegraphics[scale = 1]{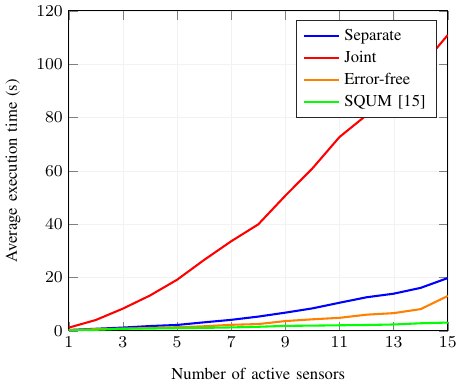} 
    \caption{Average execution time versus number of active sensors $N$.}
    \label{fig:10}
	\includegraphics[scale = 1]{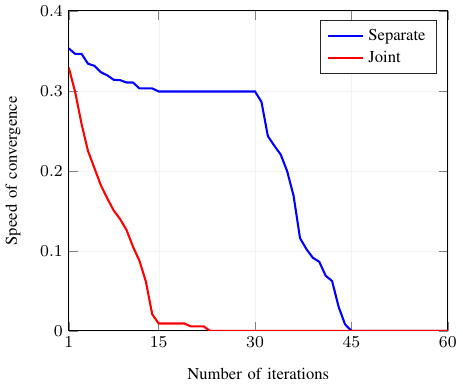} 
    \caption{Speed of convergence versus number of iterations.}
    \label{fig:11}
\end{multicols}
\end{figure*}

Following the previous discussion, in Fig.~\ref{fig:8} ($\varphi = 0.9$) and Fig.~\ref{fig:9} ($\varphi = 0.99$) we have included some additional comparisons to highlight the performance of our approach. On the one hand, we have considered a selection strategy, denoted by (E), in which the number of quantization bits (or MCS indexes) is chosen to ensure an (almost) error-free communication, i.e., $\textrm{PER} \to 0$. This way, we reduce the problem complexity since we only optimize the selection matrix $\bm{V}$. However, reliable communication might be difficult to ensure in scenarios with poor channel conditions, which usually require strong codeword protections. Also, it is important to recall that this condition translates into a low number of information bits and, thus, into high quantization errors (which increase the overall MSE). Additionally, considering the error-free policy, in both figures, we also illustrate the results obtained with the selection strategy proposed in \cite{Zha19}, which is based on successive quadratic upper-bound minimization (SQUM). As we can observe, this approach shows a behavior similar to (E), and, as before, the MSE decreases with the correlation degree $\varphi$. Nevertheless, both strategies perform poorly when compared to (S) or (J).

On the other hand, instead of the random selection, we can also consider the approach where sensors are chosen according to their channel gain. This is included in Fig.~\ref{fig:8} and Fig.~\ref{fig:9}. To avoid redundancy, only the cases of $4$-bit and $8$-bit are shown. We can see that this strategy performs well only when the correlation is high (i.e., $\varphi = 0.99$). The reason is that choosing the devices based on their channel does not consider the quality and the correlation of the information that they measure, which ultimately conditions the estimation accuracy.

As discussed in Section~\ref{sec:4}, we will also compare the computational complexity of both approaches (S and J). To do so, we will show the average execution time required to optimize the selected sensors and the number of quantization bits. The implementation of both strategies is performed with MATLAB and the resulting execution time is illustrated in Fig.~\ref{fig:10}. Besides, in Fig.~\ref{fig:11}, we depict the corresponding speed of convergence (i.e., the difference w.r.t. the final NMSE) versus the number of iterations. As we can see, the separate approach requires more iterations ($\bm{V}$ and $\bm{n}$ are found separately). However, despite the higher speed of convergence, the joint strategy needs more execution time due to the higher computational complexity (cf. Section~\ref{sec:4.2}). Thus, since the performance of the separate optimization reaches the joint approach in almost all setups\footnote{The reason behind this behavior is that, in the scenario under study, the impact of the quantization bits on the MSE is smaller than the device selection since removing information (i.e., silencing a certain sensor) degrades more the estimation accuracy than reducing the quantization precision. Therefore, both approaches can distinguish the optimal devices (those with better channel conditions and good observations) independently of their quantization level.}, the separate approach results in a better strategy. In fact, given the minor improvement achieved with the joint optimization, its formulation can be used to justify the usefulness of the separate approach. Additionally, in Fig.~\ref{fig:10}, we also included the average execution time of the aforementioned error-free strategy and the one obtained with the approach in \cite{Zha19}. As we would expect, both methods require similar but smaller values than those of the separate and joint curves (which is reasonable given that the quantization bits are not optimized).

\begin{figure*}[t]
\begin{multicols}{2}
    \centering 
    \includegraphics[scale = 1]{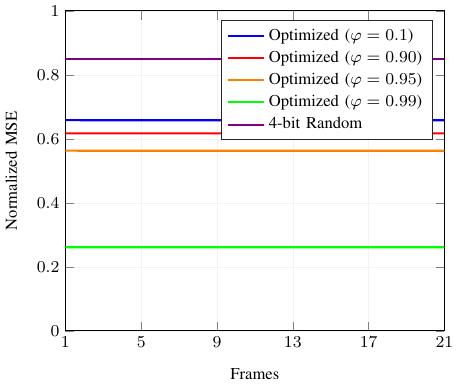} 
    \caption{Evolution of NMSE $\varepsilon(t) / \textrm{tr}(\bm{C}_{\bm{\theta}}(t))$ ($\psi = 0.1$).}
    \label{fig:12}
	\includegraphics[scale = 1]{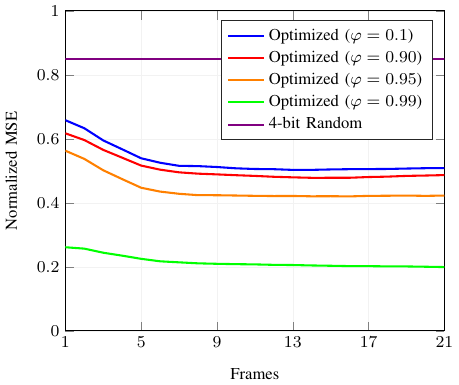}
    \caption{Evolution of NMSE $\varepsilon(t) / \textrm{tr}(\bm{C}_{\bm{\theta}}(t))$ ($\psi = 0.9$).}
    \label{fig:13}
\end{multicols}
\begin{multicols}{2}    
    \centering 
    \includegraphics[scale = 1]{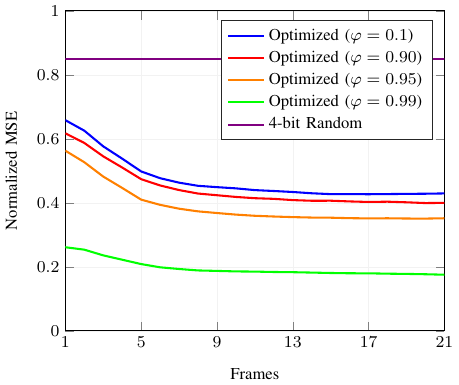}
    \caption{Evolution of NMSE $\varepsilon(t) / \textrm{tr}(\bm{C}_{\bm{\theta}}(t))$  ($\psi = 0.95$).}
    \label{fig:14}
	\includegraphics[scale = 1]{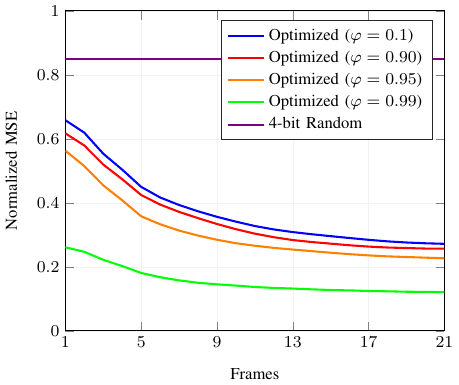}
    \caption{Evolution of NMSE $\varepsilon(t) / \textrm{tr}(\bm{C}_{\bm{\theta}}(t))$ ($\psi = 0.99$).}
    \label{fig:15}
\end{multicols}
\end{figure*}

Finally, regarding the system with memory, we consider different degrees of temporal correlation, namely $\psi = 0.1$, $\psi = 0.9$, $\psi = 0.95$, and $\psi = 0.99$. The different cases are depicted in Fig.~\ref{fig:12}, Fig.~\ref{fig:13}, Fig.~\ref{fig:14}, and Fig.~\ref{fig:15}, respectively. In each plot, we consider the previous values for $\varphi$, $N/M = 0.2$, and $M = 30$. Also, for the sake of clarity in the explanation, only the case of $4$-bit (fixed) quantization and $\varphi = 0.9$ is included as a reference. In addition, following the discussion of the memoryless case, only the separate optimization is shown.

As we can see, leveraging the temporal correlation can also help to reduce the MSE substantially. Obviously, this behavior is more notorious when larger coefficients $\varphi$ and $\psi$ are used. However, even in the case of low spatial correlation (i.e., $\varphi = 0.1$), a large temporal correlation still improves the estimation performance. This can be easily seen in Fig.~\ref{fig:13}, Fig.~\ref{fig:14}, or Fig.~\ref{fig:15}, where the MSE decreases more rapidly with high values of $\psi$. In fact, note that when $\psi$ is very low (e.g., $0.1$), the MSE barely decreases in time (even if $\varphi \to 1$). For instance, in Fig.~\ref{fig:12} we can see that the MSE is almost constant over the different frames. Overall, compared to the $4$-bit case, our approach yields larger optimization gains when the data is highly correlated.

\subsection{Sensor Selection and Distributed Quantization with Real Data} \label{sec:6.5}
Following the discussion in Section~\ref{sec:6.4}, we will study the performance of our estimation when the measurements are obtained from the Intel database \cite{Bod04} as described in Section~\ref{sec:6.2}. However, to avoid redundancy, we will omit the analysis of the accuracy of the upper bound $\bar{\varepsilon}_K$ since the previous reasoning is still valid (cf. Fig.~\ref{fig:3}).

\begin{figure*}[t]
\begin{multicols}{2}
    \centering 
    \includegraphics[scale = 1]{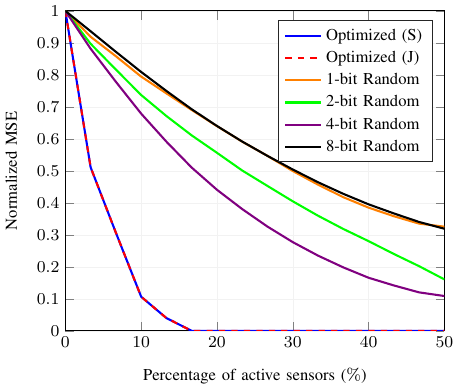}
    \caption{Normalized MSE $\varepsilon / \textrm{tr}(\bm{C}_{\bm{\theta}})$ versus percentage of active sensors $N/M \cdot 100$ (Intel database).}
    \label{fig:16}
    \includegraphics[scale = 1]{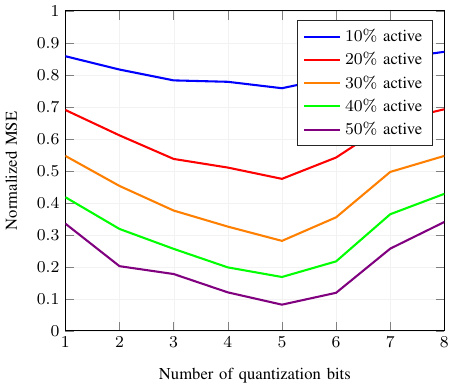}
    \caption{Normalized MSE $\varepsilon / \textrm{tr}(\bm{C}_{\bm{\theta}})$ versus number of quantization bits $n_i$ (Intel database).}
    \label{fig:17}
\end{multicols}
\begin{multicols}{2}    
    \includegraphics[scale = 1]{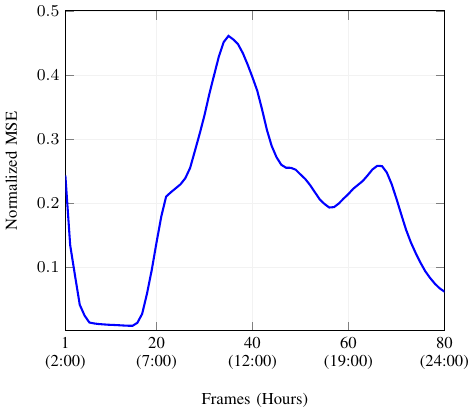}    
    \caption{Evolution of normalized MSE $\varepsilon(t) / \textrm{tr}(\bm{C}_{\bm{\theta}} (t))$ (Intel database).}
    \label{fig:18}    
    \includegraphics[scale = 1]{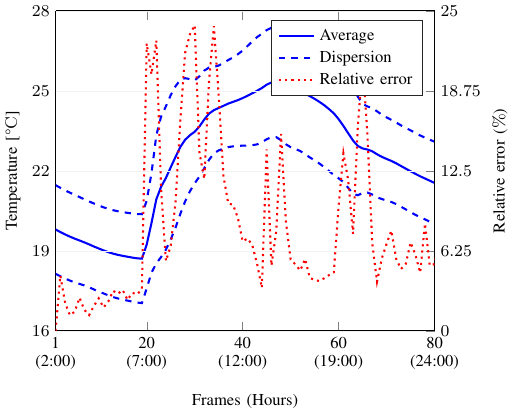}
    \caption{Average temperature $\bar{\theta}(t)$, dispersion $\bar{\theta}(t) \pm \hat{\theta}(t)$ and relative error $\xi(t)$ over the day (Intel database).}
    \label{fig:19}
\end{multicols}
\end{figure*}

Thereby, considering the memoryless case, we will show the normalized MSE w.r.t. the percentage of active sensors. This is depicted in Fig.~\ref{fig:16}. As before, the performance improves when more sensors are allowed to transmit. In fact, we can observe the same behavior as that with a high correlation factor (cf. Fig.~\ref{fig:7}): a small error is attained with few sensors. Thus, since almost all measurements are identical, our approach retrieves most of the information with $10 \%$ of active sensors (equivalent to a power reduction of $90 \%$). Contrarily, although the random selection does not yield poor results, it needs a considerably larger number of active sensors to attain the same performance (around $50\%$ are needed to obtain the same error as the optimized case with $10\%$ of active sensors, i.e., a normalized MSE of $0.1$). 

Additionally, for the sake of clarity in the explanation, in Fig.~\ref{fig:17} we present the MSE w.r.t. the number of quantization bits and different percentages of active sensors. A similar behavior can be observed when the number of bits is small (e.g., $1$ bit) or high (e.g., $8$ bits). This is because too few bits imply a low precision (or high quantization noise) for the parameter estimation and too much bits represent a small message protection (or high error probability). This reveals the need for choosing the proper number of quantization bits.

On the other hand, the evolution of the (normalized) MSE over time is depicted in Fig.~\ref{fig:18} for $3.33 \%$ of active sensors. The frames here indicate consecutive time periods of the day. Note that, differently from before, the error does not always decrease monotonically. Instead, there are some frames where there is a huge increase. The reason behind this behavior is the dynamical modeling of $\bm{\theta}(d,t)$, presented in \eqref{eq:54}, in which we assume a common transition matrix over the days. However, this model does not capture entirely the nature of the measurements in \cite{Bod04}, i.e., there are some instants where the evolution is not common over the days. 

To illustrate this, in Fig.~\ref{fig:19} we show the sensors average temperature during the day $\bar{\theta} (t)$, the standard deviation $\hat{\theta}(t)$ over the days, and the relative error in the second-order statistics of the dynamical model $\xi(t)$:
\begin{equation}
    \bar{\theta}(t) = \frac{1}{D} \sum_{d = 1}^{D} \bm{1}_M^{\textrm{T}} \bm{\theta}(d,t)/M,
    \label{eq:58}
\end{equation}
\begin{equation}
\hat{\theta}(t) = \sqrt{\frac{1}{D} \sum_{d = 1}^{D} \left(\bm{1}_M^{\textrm{T}} \bm{\theta}(d,t)/M\right)^2 - \bar{\theta}(t)^2},
\label{eq:59}
\end{equation}
\begin{equation}
\xi(t) = \|\underbrace{\alpha(t)^2\bm{C}_{\bm{\theta}}(t-1) + \bm{C}_{\bm{\nu}}(t)}_{\textrm{model}} - \underbrace{\bm{C}_{\bm{\theta}}(t)}_{\textrm{reality}} \|_{\textrm{F}} / \| \bm{C}_{\bm{\theta}}(t) \|_{\textrm{F}}.
\label{eq:60}
\end{equation}

Note that a rapid change in the temperature and a large dispersion (e.g., frame $20$) result in a poor dynamical modeling (i.e., a single $\alpha (t)$ per time instant does not capture properly the statistical evolution of $\bm{\theta}(d,t)$ across the days) and, thus, the estimation fails. Despite that, the proposed approach is able to adapt to these changes as the resulting MSE is reduced after the peaks. In fact, whenever the relative error is small, our estimation scheme yields a good accuracy. 

\section{Conclusions} \label{sec:7}
In this paper, we have addressed the problem of estimating a set of measured parameters in an UL mMTC network. Considering a scenario where a group of sensors send the spatio-temporally correlated observations to a CN, we have derived an estimation strategy based on the MMSE estimate and Kalman filters that takes into consideration the energy restrictions of these devices in practical systems. Given that communication errors may compromise the estimation performance, we have averaged the MSE over the different decoding probabilities and proposed a device selection scheme and quantization approach that minimize the resulting MSE. This way, since the number of active sensors and the information to be transmitted are significantly reduced, we have been able to decrease the data traffic and improve the power consumption. 

Our approach has been evaluated in several setups with synthetic and real data. Simulation results have shown that in the case of synthetically generated data, our scheme can reduce the power consumption by $50 \%$ (i.e., number of silent sensors) without a significant increase in the MSE. This behavior is more notorious in the case of real data, where the spatial correlation is significantly higher. With only $10 \%$ of active sensors, we attain a performance in normalized MSE of $0.1$. In both cases, a better performance is observed when including temporal correlation.

There are some future research lines that can be considered to
extend the work presented in this paper. Firstly, the study of non-uniform scalar quantizers could be useful to improve the distribution of quantization (information) bits. Secondly, for a more faithful representation of practical systems, multiple access channels with non-orthogonal resources could be investigated. Finally, vector quantization techniques could be used in two scenarios: i) setups where each individual sensor measures multiple phenomena at the same time; and ii) multi-hop networks where the CN quantizes the decoded measurements before retransmission.

\bibliographystyle{IEEEtran}
\bibliography{IEEEabrv,References}

\end{document}